\documentclass{jaa}
\usepackage{natbib}

\usepackage{graphicx}
\usepackage{xcolor}
\usepackage{hyperref}
\definecolor{darkgreen}{rgb}{0.0, 0.5, 0.0}
\definecolor{britishracinggreen}{rgb}{0.0, 0.26, 0.15}
\definecolor{burntorange}{rgb}{0.8, 0.33, 0.0}
\hypersetup{colorlinks, citecolor = burntorange, urlcolor =cyan}

\begin{document}\sloppy

\title{Absolute Time Calibration of LAXPC aboard AstroSat}


\author{Avishek Basu\textsuperscript{1,3}, Dipankar Bhattacharya\textsuperscript{2} and Bhal Chandra Joshi\textsuperscript{3}}
\affilOne{\textsuperscript{1}Jodrell Bank Centre for Astrophysics, University of Manchester, Oxford Road, Manchester, M13 9PL, UK\\}
\affilTwo{\textsuperscript{2}Inter University Centre for Astronomy and Astrophysics, Post Bag 4, Pune 411007, India\\}
\affilThree{\textsuperscript{3}National Centre for Radio Astrophysics, Tata Institute of Fundamental Research, Ganeshkhind, Pune 411007, ~~~India}


\twocolumn[{

\maketitle

\corres{avishek.basu@manchester.ac.uk}


\begin{abstract}
The AstroSat mission carries several high-energy detectors meant for fast timing studies of cosmic sources.  In order to carry out high precision multi-wavelength timing studies, it is essential to calibrate the absolute time stamps of these instruments to the best possible accuracy.  We present here the absolute time calibration of the AstroSat LAXPC instrument, utilising the broad-band electromagnetic emission from the Crab Pulsar to cross calibrate against Fermi-LAT and ground based radio observatories Giant Metrewave Radio Telescope (GMRT) and the Ooty Radio Telescope (ORT). Using the techniques of pulsar timing, we determine the fixed timing offsets of LAXPC with respect to these different instruments and also compare the offsets with those of another AstroSat instrument, CZTI.
\end{abstract}


}]


\doinum{12.3456/s78910-011-012-3}
\artcitid{\#\#\#\#}
\volnum{000}
\year{0000}
\pgrange{1--}
\setcounter{page}{1}
\lp{1}

\section{Introduction}\label{intro}
Multiwavelength observations are key to unravelling the physical processes ongoing in a variety of astrophysical sources. Such observations commonly involve multiple instruments situated at different locations.  The target sources being studied often have time variable intensity and spectrum, so to characterise their properties it is essential to synchronise the intrinsic time stamps at various observatories being used.  Even if the reference time information is obtained from a common source like the Global Positioning System, there could be internal delays in data processing electronics which need to be measured to achieve the necessary synchronisation.  In this paper we report the result of our attempt to calibrate the timing offset of the AstroSat LAXPC instrument with respect to ground based Indian radio observatories, namely the Ooty Radio Telescope (ORT) and the Giant Metrewave Radio Telescope (GMRT), the Large Area Telescope (LAT) instrument aboard the Fermi gamma-ray observatory. Results of a similar timing offset calibration experiment for another AstroSat instrument, the Cadmium Zinc Telluride Imager (CZTI), have been reported in \cite{Basu+2018}, and are also included here for comparison.

This work is motivated by our ongoing effort to characterise the multi-wavelength properties of the Giant Radio Pulses (GRP) emitted by radio pulsars from time to time.  In order to phase align these pulses between radio and X-ray bands an accurate alignment of the time stamps across the instruments is required. We therefore begin by measuring the timing offsets between the instruments involved in our experiment, the results of which we report here. We aim to obtain the offsets precise enough to track every pulse in the radio and X-ray bands unambiguously. The GRP arrives randomly within the $\sim3$~ms wide on-pulse window in the average pulse profile \citep{heiles1970GP,ATNFcatalogue}. Our desirable uncertainty on the offset measurement is one-tenth of the pulse width, i.e.\ less than 300 $\mu$s.
\\

Our aim can be achieved by timing and monitoring pulsars over a long span of time. Since our experiment involves telescopes operational at low-frequency radio wavelengths to $\gamma-$rays, we have chosen the Crab pulsar (PSR J0534+2200) as our target source. The Crab pulsar emits pulsed emission across the whole electromagnetic spectrum from radio to very high-energy $\gamma$-rays. At L-band (around 1.4 GHz) the Crab pulsar has two distinct components in its profile: the main-pulse (MP) and a relatively weaker inter-pulse (IP). The components are highly aligned across the spectrum except for some intrinsic emission delay. The MP at high energies leads the radio main pulse by 241 $\pm$ 29 $\mu$s \citep[$>$ 30 MeV ;][]{kuiper2003absolute}, 344 $\pm$ 40 $\mu$s  \citep[2-30 keV ;][]{rots2004absolute}, (280 $\pm$ 40 $\mu$s) \citep{kuiper2003absolute}, 235$\pm 33 \mu$s \citep[10-600 keV;][]{Terada+2008} and 275$\pm15 \mu$s \citep[20-100 keV;][]{Molkov_2009}. It is essential to account for these intrinsic delays while computing the time of arrival (TOA) of the pulses (discussed in Section \ref{czti-lax-analysis}).
\\

The results obtained from our experiment presented in this paper would allow us to time-align the Radio and X-ray time series data, enabling us to search for X-ray photon count enhancements coincident with the GRP. 
In  Section \ref{Instruments} we discuss the instruments used for the experiment, in  Section \ref{Method} we discuss the methodology adopted to measure the timing offsets and finally conclude the paper by presenting the results in Section \ref{results}

\section{Instruments and Observations}\label{Instruments}

We performed multi-epoch, multi-frequency observations using Indian facilities like the Giant Metrewave Radio Telescope (GMRT), the Ooty Radio Telescope (ORT) operational at radio wavelength, and two payloads aboard AstroSat, the Cadmium Zinc Telluride Imager (CZTI), and the Large Area Proportional Counter (LAXPC). We have also made use of the publicly available data from \textit{Fermi-}LAT operational at  high-energy $\gamma$-rays ($\sim$ 20 MeV $-$ 300 GeV).

\subsection{AstroSat}\label{astrosat}
AstroSat, India's first space based observatory was launched in October 2015 with five payloads on board \citep{singh2014astrosat}. These are  
the Cadmium Zinc Telluride Imager  \citep[\textit{CZTI};][]    {bhalerao2016cadmium}, the   
Large Area X-ray Proportional Counter  
\citep[\textit{LAXPC};][]{yadav2016large}, the
Soft X-ray Telescope \cite[\textit{SXT};][]{singh2016orbit}, the Ultra Violet 
Imaging Telescope  \citep[\textit{UVIT};][]{hutchings2014uvit} and the Scanning Sky Monitor (\textit{SSM}) allowing observations covering a wide frequency range from 1300 \AA\ to 380 keV.

\subsubsection{LAXPC}\label{laxpc}
The LAXPC is the prime X-ray detector on AstroSat, operating in the energy range 3--80~keV. LAXPC consists of three proportional counter units filled with primarily Xenon gas at a pressure of about 2 atmosphere, presenting a combined effective area of $\sim 6000$~cm$^2$ at energies below 20~keV, declining to $\sim 2500$~cm$^2$ at 80~keV \citep{antia2017laxpccal}. This large effective area, along with high timing resolution (10~$\mu$s) makes this an excellent instrument for X-ray timing studies, including those of pulsars. A collimator restricts the Field of View (FOV) of LAXPC to approximately $1^{\circ}\times 1^{\circ}$. LAXPC detectors record event mode data, with each event tagged with an instrument time stamp derived from a System Time Base Generator (STBG) driven by a temperature controlled crystal oscillator.  Once every 16 seconds, a synchronising pulse is sent to all AstroSat instruments including the on-board Spacecraft Positioning System (SPS), which provides time in UTC based on the Global Positioning System.  All the instruments record their current time stamp at the arrival of the synchronising pulse and these values are collected in a Time Correlation Table which is used in offline analysis to convert, by interpolation, the instrument time stamps assigned to the recorded events to UTC time stamps \citep{bhattacharya2017}.

\subsubsection{CZTI}\label{czti}
The Cadmium Zinc Telluride Imager extends the high-energy coverage of AstroSat to $\sim 380$~keV, starting from $\sim 20$~keV.  It consists of a solid state, pixellated CZT detector array of total geometric area $\sim 976$~cm$^2$, with a collimator and a Coded Aperture Mask situated above it.  The Coded Mask and the collimator provide a $4.6^{\circ}\times 4.6^{\circ}$ imaging Field of View at energies below $\sim 100$~keV and gradually become transparent at higher energies. The CZTI records photon events time stamped at 20~$\mu$s resolution by its internal clock.  These time stamps are converted to UTC time stamps during offline analysis in the same manner as for the LAXPC instrument.  The CZTI has carried out extensive studies of the Crab pulsar at high energies, including that of its polarization \citep{vadawale2018crabpol}.

\subsubsection{Orbit determination}\label{as1orbit}
Comparing time stamps across observatories requires the arrival times to be referred to a common reference system, for which we adopt the Solar System Barycentre. The event time stamps recorded by AstroSat are referred to the corresponding Barycentric arrival times, using the knowledge of the orbit of the satellite. The orbital position and velocity of AstroSat are measured by an on-board 10-channel Spacecraft Positioning System (SPS) unit that operates on signals received from the Global Positioning System (GPS) satellites. These measurements are regularly calibrated against those obtained by ranging from the AstroSat ground station.  The housekeeping data stream of AstroSat provides the orbital position values sampled every 128 milliseconds, with an accuracy of better than 5 metres.  The error budget in the barycentric correction arising from the uncertainties in the orbital position is thus limited to less than 0.017 $\mu$s. We have ignored this contribution in the reported uncertainties in our final results, which are much larger.

\subsection{\textit{Fermi-}LAT} \label{Fermilat}
The \textit{Fermi-LAT} is a high-energy $\gamma-$ray telescope sensitive to the photons with the energy from below 20 MeV to more than 300 GeV \citep{Atwood+2009}. It monitors $\gamma-$ray pulsars with a cadence of one-sixth of its duty-cycle. Individual photon events are recorded with a time resolution better than 1 $\mu$s \citep{smith2008}. We use data of the Crab pulsar retrieved from the public archive\footnote{https://fermi.gsfc.nasa.gov/cgibin/ssc/LAT/LATDataQuery.cgi} of the Fermi mission.

\subsection{The Giant Metrewave Radio Telescope (GMRT)}\label{GMRT}
The GMRT \citep{swarup1991giant} is an ``Y''-shaped interferometer with thirty, 45-m steerable dishes operational at low-frequency radio-wavelengths. Fourteen antennas are arranged in a compact array within a radius of 1 km, the remaining antennas are arranged in three arms. The observations were carried out by combining all 14 antennas and the first arm antennas in a tied array with an overall gain of 3.5K/Jy. The Crab pulsar was observed using the GMRT at seven different epochs (shown with green markers in Figure \ref{with Timing noise}). The typical observation duration was 1-2 hours. The time-series raw voltage data acquired at 1390 MHz with 16 MHz bandwidth from every antenna were Fourier Transformed to obtain 256 channels voltage data. The instrumental phase lags among the antennas were determined by observing the point source 3C147, which were then compensated in the Fourier domain and added coherently. Further analysis was done offline described in Section \ref{Method}

\subsection{The Ooty Radio Telescope (ORT)}\label{ORT}
The ORT is a 30 m wide offset parabolic cylindrical antenna in the east-west direction. It is 530 m long in the north-south direction sensitive to a single polarisation and operational at 334.5 MHz \citep{swarup1971large}. The gain of the telescope is 3.3 K/Jy and the system temperature is 150 K. The pulsar observations back-end at ORT is called as PONDER \citep{pondernaidu}, which starts recording the data on arrival of the rising edge of the minute pulse obtained from the GPS system. PONDER performs coherent de-dispersion in real-time and produce time-stamped folded pulse-profiles in ASCII format. PSR J0534+2200 was observed for 15 minutes daily as a part of a larger pulsar monitoring program \citep{kjm+18} and the high cadence pulsar glitch monitoring program at the ORT \citep{Basu+2019}. In this paper we have used the data from September 01, 2015 (MJD 57226) to January 14, 2017 (MJD 57767)

\section{Methodology}\label{Method}
The method for the absolute time calibration relies on the technique of pulsar timing. The technique of pulsar timing \citep{Edwards+2006} compares the observed TOAs with the predicted TOAs obtained from a simple rotation model of the pulsars. We perform the analysis in multiple steps in an iterative manner until the best solution is achieved.

\subsection{ORT-Analysis}\label{ort-step1}
As mentioned earlier in Section \ref{ORT}, coherently de-dispersed time-stamped profiles are obtained from the ORT. The TOA of a pulse was computed using the software package PSRCHIVE \citep{hsm04} from every profile by cross-correlating with a noise-free template of the pulse profile in the frequency domain described in \cite{taylor1992pulsar}. The typical timing accuracy\footnote{We refer the median of the TOA errors as the ``typical timing accuracy''} at ORT is 318 $\mu$s. The noise-free template was created in PSRCHIVE\footnote{http://psrchive.sourceforge.net/} by fitting an optimal number of Gaussian waveform to a high S/N observed pulse profile. The TOAs obtained from our high cadence observations were then used to create a phase connected solution. Such a high cadence is especially important for the Crab pulsar to obtain a reliable phase connected solution because of the strong timing noise. The timing noise is observed as systematic wandering in the TOA residuals after removal of the standard spin-down model \citep{Cordes_Helfand1980}. The TOAs were further segmented with a time span of a month to produce the monthly ephemeris from our data by fitting the spin-down model in the high precision pulsar timing package TEMPO2\footnote{https://bitbucket.org/psrsoft/tempo2/src/master/} \citep{hobbs2006tempo2}. The monthly ephemeris with precise rotation parameters was used to re-fold the time-series data to obtain the precise TOAs. The Crab nebula provides a strong scattering screen to the radio waves emitted from the pulsar. The effect of scatter broadening is pronounced at 334.5 MHz and can contribute to the timing residuals as a systematic. Hence, in case of the Crab pulsar, it is difficult to de-couple the effect of timing noise from the scatter broadening, which in our analysis has been taken care by using the publicly available data from \textit{Fermi}-LAT.
\subsection{\textit{Fermi}-LAT Analysis}\label{Fermi-analysis} 
The $\gamma$-ray pulse profiles are free from the propagation effects, therefore the \textit{Fermi}-LAT \citep{fermi-LAT} archival data\footnote{https://fermi.gsfc.nasa.gov/cgibin/ssc/LAT/LATDataQuery.cgi}  were used to model the timing noise which is a frequency-independent phenomenon. We use all the events in the energy range 0.1 to 300 GeV within a radius of 3$^\circ$ around PSR J0534+2200. The event data were split using Fermi science tool\footnote{https://fermi.gsfc.nasa.gov/ssc/data/analysis/scitools/overview.html} into smaller event data each of 7 days duration. These time stamps were referred to the solar system barycentre (SSB) adopting JPL planetary ephemeris DE200 and folded using the Fermi-plugin \citep{rkp+11} of TEMPO2 using the ephemeris obtained from the ORT timing solutions. The standard template was constructed in a similar manner as explained in Section \ref{ort-step1} However, to account for the intrinsic delay between the radio and the $\gamma-$ray pulse profile the templates were aligned with an appropriate shift mentioned in the Section \ref{intro} The TOAs were computed from every pulse profile by cross-correlating the standard template. The timing accuracy was 309 $\mu$s. The timing analysis was performed using TEMPO2. The timing noise at this band was modelled with the combination of eight sine waves to obtain the white timing residuals using the FITWAVES tool in TEMPO2.

\subsection{Re-analysis of the ORT data}\label{ort-step2}
The timing solution with modelled timing noise from the \textit{Fermi}-LAT TOAs was applied on the ORT TOAs. At this stage, the TOAs affected from the scatter broadening were removed from our analysis and monthly ephemeris were re-generated. Hence, the rotation parameters were obtained which are free from timing noise and the scatter broadening effects. We refer to this timing solution as the ``iteration-2'' solution. The iteration-2 solutions were used to re-fold the ORT time-series data and produce TOAs following similar steps as explained in Section \ref{ort-step1}.

\subsection{GMRT Analysis}\label{GMRT-analysis}
The raw voltage data obtained from the GMRT were also coherently de-dispersed offline with our pipeline discussed in \cite{pondernaidu}. The values of the dispersion measure (DM) were taken from the Jodrell Bank monthly ephemeris\footnote{http://www.jb.man.ac.uk/pulsar/crab.html} \citep{lyne199323} nearest to the epoch of observations. The de-dispersed time-series were further folded using the iteration-2 monthly ephemeris obtained from the ORT data. The GMRT offline analysis also supplies with de-dispersed 64 channel sub-banded data with 32 sub-integrations. The standard template was constructed from the highest S/N observed profile following the same method explained in Section \ref{ort-step1}. The template of the pulse profile from the GMRT data was aligned with the ORT templates and then the TOAs were computed following the method discussed in Section \ref{ort-step1}. The timing accuracy obtained at GMRT is 162 $\mu$s. The dynamic pulsar wind within nebular filaments leads to the variation in the electron density along the line of sight, which results in time variation of the DM. The typical variation in DM is of the order of 0.01 pc cm$^{-3}$, which incorporates a change in time of arrival by 21 $\mu$s and 370 $\mu$s at 1390 MHz and 334.5 MHz respectively. Therefore it is essential to correct for DM variations to obtain reliable and precise estimates of the offsets.
The TOAs computed from the ORT data in Section \ref{ort-step2} and from the GMRT data were used to measure the DM at different epochs. The fixed offset between
the data acquisition pipelines at the ORT and GMRT were known from previous
measurements \citep{Surnis+2018}. Hence, the DM was estimated after accounting for this delay between GMRT and ORT using the JUMP parameter in TEMPO2. It may be noted that only 8 observations were nearly simultaneous between ORT and GMRT. Therefore,
8 different estimates of DM at 8 different epochs were obtained \citep[Figure 4 of][]{Basu+2018}. The DM was further used to perform the offline coherent de-dispersion to obtain the de-dispersed time series data, which were folded using the iteration-2 timing solutions. The TOAs from the GMRT data were finally produced by following the methods described above.

\subsection{AstroSat-CZTI and LAXPC Analysis} \label{czti-lax-analysis}
The CZTI has four detectors arranged in four quadrants. There are no relative offsets between individual quadrants. Hence data from all the four quadrants were combined and the time tags of the photons were converted to SSB adopting the JPL planetary ephemeris DE200 and using the satellite position in the code \textit{as1bary}\footnote{http://astrosat-ssc.iucaa.in/?q=data\_and\_analysis}.
The barycentre recorded events were folded to construct the pulse profile using iteration-2 timing solution obtained in Section \ref{ort-step2} using our own codes. Further, the standard template was created following the same method as described in Section \ref{ort-step1}. In case of LAXPC, the event files were created by combining the data from three consecutive orbits, which were then barycentred using the \textit{as1bary} and folded using the iteration-2 timing solution obtained in the Section \ref{ort-step2}. The standard templates were created following a similar method as mentioned above. The templates obtained from the CZTI and the LAXPC were appropriately shifted with respect to those obtained from the GMRT, ORT and \textit{Fermi-}LAT to take into account the intrinsic energy-dependent emission delays. Finally, the TOAs were computed for the CZTI profiles and the LAXPC profiles using the templates thus constructed. This method of incorporating the known energy-dependent intrinsic emission delays in the template construction allows us to find the true clock offsets between two instruments directly from the TOA differences between them.
\begin{figure}
    \centering
    \includegraphics[scale=0.43]{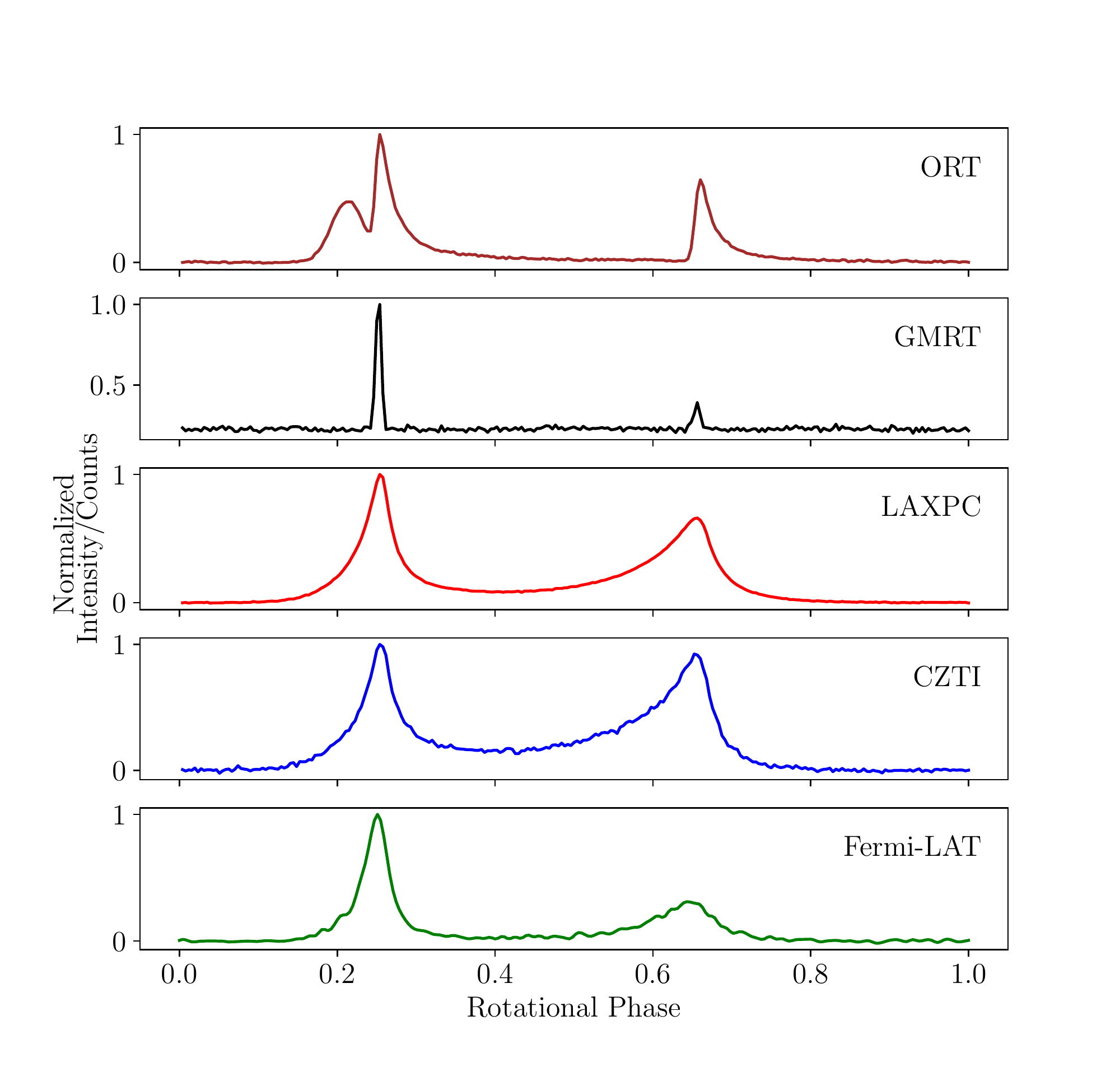}
    \caption{The multi-wavelength pulse profiles of the Crab pulsar. The pulse profiles were aligned after correcting the clock-offsets between the telescopes. }
    \label{pulseprofile}
\end{figure}

\subsection{Offset Measurements}
\begin{figure*}
    \centering
    \includegraphics[scale=0.8
    ]{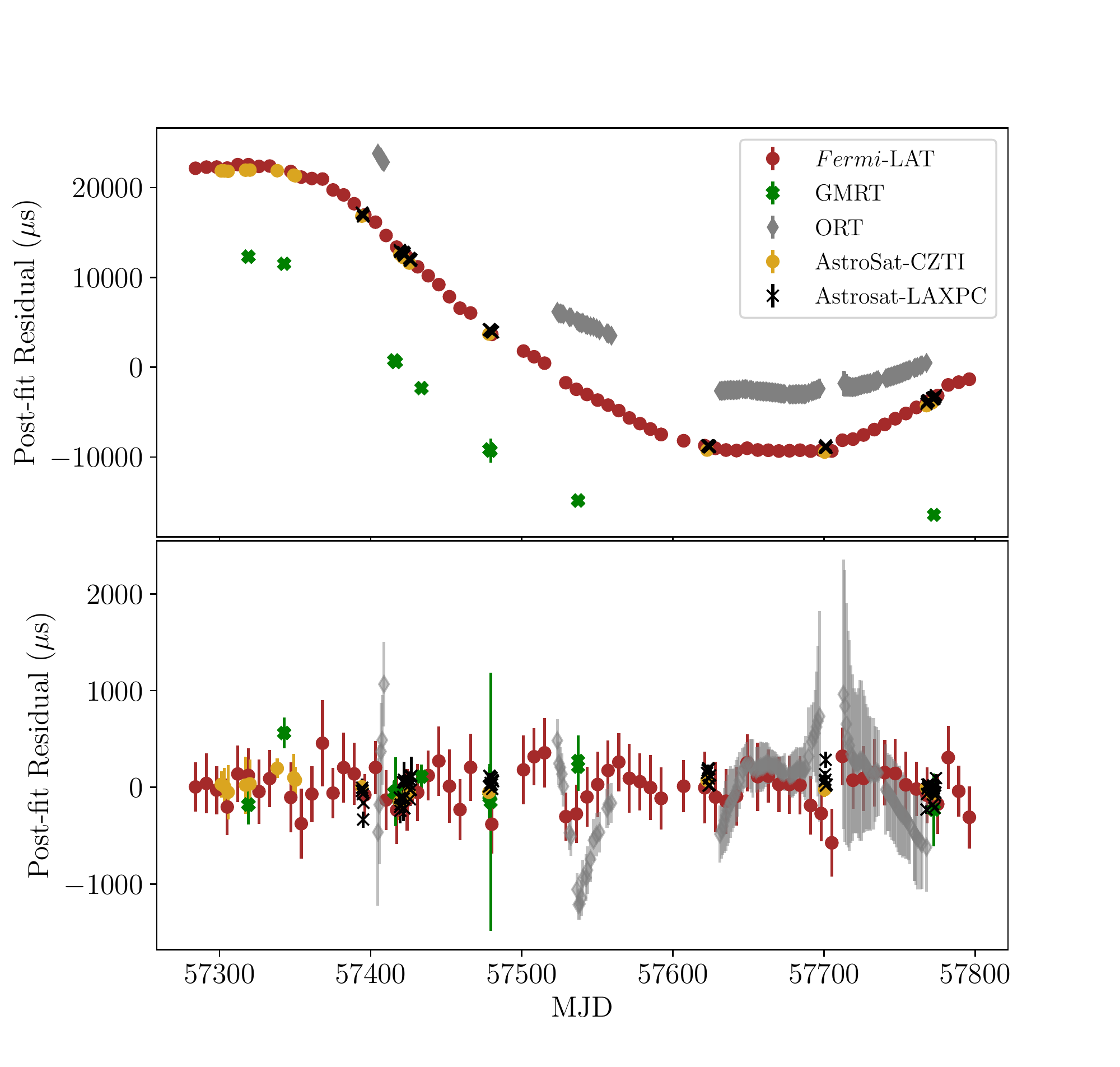}
    \caption{The timing residuals from different telescopes. \textit{Upper panel:} The green, grey, red, black and yellow data points represent the TOAs obtained from GMRT, ORT, \textit{Fermi}-LAT,
    LAXPC and CZTI observations, respectively. The systematic pattern is called as timing noise and the parallel tracks indicate the timing offset incorporated by different telescopes back end as explained in the text. \textit{Lower Panel:} The TOA residuals from all the observations after modelling the timing noise, the effect of dispersion measure and clock offsets (detailed description is given in the text).}
    \label{with Timing noise}
\end{figure*}

The TOAs from all the telescopes were collated together to compute the offsets between the different telescopes. These TOAs were analysed using 
a timing model obtained by merging the pulsar rotation model, which gave the phase connected solution,  with a constrained DMMODEL in TEMPO2. The DMMODEL is obtained from the DM time series, fitted from simultaneous observations at the ORT and the GMRT using the procedure described in Section \ref{GMRT-analysis}. Inclusion of DMMODEL terms in the timing model accounts for the DM offsets from the chosen reference DM.  The timing residuals obtained by applying this timing model to TOAs from all the telescopes have been shown in the upper panel of Figure \ref{with Timing noise}. The systematic trend as a function of time in these residuals for all telescopes is due to the timing noise of the pulsar.  The residuals represent the difference between the predicted and the observed TOAs. As the TOA from each telescope additionally consists of a clock offset which is fixed with respect to the observation epoch, the residuals of a pair of telescopes are seen as parallel tracks in the diagram. 
Thus, fitting a constant difference to residuals of a pair of telescope in Figure \ref{with Timing noise} determines the timing offsets between the telescopes.
The timing model, described above, was merged with the timing noise model obtained in Section \ref{Fermi-analysis}. This reference timing solution after accounting for the timing noise and the DM variation is given in Table \ref{timsol}. Finally, we use the JUMP feature of TEMPO2 to measure the offsets between the telescopes discussed in Section \ref{results}

\begin{table}[h]
\begin{tabular}{ll}
\hline \hline
Pulsar parameter     & Value         \\ \hline \hline
RAJ  (hh:mm:ss)      &  05:34:31.973 \\
DECJ (dd:mm:ss)      &  +22:00:52.06 \\
F0   (Hz)            &   29.6607409(4) \\
F1   (Hz s$^{-1}$)    &   $-$3.6937842(9) E$-$10 \\
F2   (Hz s$^{-2}$)    &   1.1905(3) E$-$20  \\
PEPOCH (MJD)         &  57311.000000136 \\
POSEPOCH (MJD)       &  40675 \\
DMEPOCH  (MJD)       &  57311.000000136 \\
DM    (pc\,cm$^{-3}$)  &  56.7957 \\
PMRA  (mas/year)     &  $-$14.7 \\
PMDEC (mas/year)     &  2 \\
WAVE\_OM  (year$^{-1}$)  &   0.0054325986245627 \\
Solar system planetary \\ ephemeris & DE200\\
WAVEEPOCH (MJD)  &  57311.000000136 \\ 
START (MJD) & 57278 \\
FINISH (MJD) & 58026 \\\hline
\end{tabular}
\\
\caption{Table presents the reference timing solution of the Crab pulsar after considering the effect of DM variation and the timing noise.}
\label{timsol}
\end{table}

\section{Results}\label{results}
TEMPO2 allows one to measure the offsets between different telescopes using the JUMP feature. Utilising this, we estimate the offset between the GMRT and CZTI to be $-4716\pm50\, \mu$s and that between GMRT and LAXPC to be $-5689\pm23\, \mu$s. The measured offset between the GMRT and ORT is $-29639\pm50\, \mu$s, between GMRT and \textit{Fermi-}LAT is $-5368\pm56\, \mu$s. These clock offsets have been further tabulated in Table \ref{tab} {The phase aligned pulse profile after accounting for the offsets has been presented in the Figure \ref{pulseprofile}}. In the bottom panel of Figure \ref{with Timing noise} we present the timing residuals obtained after removing the timing offsets between them. The trend-free residuals imply that all the pulsar parameters and clock offsets have been properly modelled. The clock offset between the LAXPC and the CZTI instruments aboard AstroSat is found to be 969$\pm51 \,\mu$s. The uncertainties in the offsets are obtained from those of the parameters fitted to the TOA using the JUMP function. The results presented here meet the desired accuracy (see Section \ref{intro}) for a multi-wavelength investigation of the GRP from the Crab pulsar with the instruments used in this paper.

\begin{table}[]
\label{tab}
\begin{tabular}{ll}
\hline \hline
Instrument     & Clock-offsets in $\mu$s\\ \hline \hline
AstroSat-CZTI  & -4716$\pm$50  \\ \hline
AstroSat-LAXPC & -5689$\pm$23  \\ \hline
Fermi-LAT      & -5368$\pm$56  \\ \hline
ORT            & -29639$\pm$50 \\ \hline
\end{tabular}
\caption{The table summarises the clock offsets of different telescopes given in the first column with respect to the GMRT.}
\end{table}

\section*{Acknowledgements}
We thank the anonymous referees for their valuable suggestions to improve the presentation of the paper. This   publication   made   use   of   data   from   the   Indian astronomy mission AstroSat, archived at the Indian Space Science Data Centre (ISSDC). The LAXPC data were processed at the Payload Operations Centre at TIFR Mumbai.  The CZT Imager instrument was built by a TIFR-led consortium of institutes across India, including VSSC, ISAC, IUCAA, SAC, and PRL. The Indian Space Research Organisation funded, managed and facilitated the project. We thank the staff of the Ooty Radio Telescope and the Giant Metrewave Radio Telescope for taking observations over such a large number of epochs. Both these  telescopes  are  operated  by  National  Centre  for  Radio  Astrophysics  of Tata Institute of Fundamental Research. PONDER backend, used in this work, was built with TIFR XII plan grants 12P0714 and 12P0716. BCJ acknowledges  support  for  this  work  from  DST-SERB grant EMR/2015/000515. BCJ and AB acknowledges the support of Department of Atomic Energy, Government of India, under projcet \# 12-R\&D-TFR-5.02-0200. AB acknowledges the support from the UK Science and Technology Facilities Council (STFC). Pulsar research at Jodrell Bank Centre for Astrophysics and Jodrell Bank Observatory is supported by a consolidated grant from STFC.
\vspace{-1em}


\bibliographystyle{apj}
\bibliography{reference.bib}
\end{document}